\documentclass{article}
\usepackage[margin=1in]{geometry}
\usepackage{amsmath, amssymb}
\usepackage{graphicx}
\usepackage{hyperref}

\hypersetup{
  colorlinks=true,
  linkcolor=blue,
  citecolor=blue,
  urlcolor=blue
}

\title{QubitSwap: The Informational Edge in Decentralised Exchanges}
\author{
  Oliver Tronn Scott-Simons\thanks{\texttt{oliver@invariance.com}} \\
  Chris Colman\thanks{\texttt{chris@invariance.com}} \\
  FrostByte\thanks{\texttt{frostbyte@caviarnine.com}}
}
\date{March 24, 2025}

\begin{document}

\maketitle

\begin{abstract}
Decentralised exchanges (DEXs) have transformed trading by enabling trustless, permissionless transactions, yet they face significant challenges such as impermanent loss and slippage, which undermine profitability for liquidity providers and traders. In this paper, we introduce QubitSwap, an innovative DEX model designed to tackle these issues through a hybrid approach that integrates an external oracle price with internal pool dynamics. This is achieved via a parameter \( z \), which governs the balance between these price sources, creating a flexible and adaptive pricing mechanism. Through rigorous mathematical analysis, we derive a novel reserve function and pricing model that substantially reduces impermanent loss and slippage compared to traditional DEX frameworks. Notably, our results show that as \( z \) approaches 1, slippage approaches zero, enhancing trading stability. QubitSwap marks a novel approach in DEX design, delivering a more efficient and resilient platform. This work not only advances the theoretical foundations of decentralised finance but also provides actionable solutions for the broader DeFi ecosystem.
\end{abstract}

\section{Introduction and Problem Statement}
Traditional decentralised exchanges (DEXs) like Uniswap use a constant product formula,
\[ x \cdot y = k, \]
where \( x \) and \( y \) are reserves of two assets, and \( k \) is a constant~\cite{uniswap_v2_docs}. This approach exposes liquidity providers to significant impermanent loss (IL) during price volatility.

QubitSwap introduces a hybrid model that blends the internal pool price with an external oracle price to reduce IL and improve price stability.

\section{Core Mathematical Model}
The classic automated market maker (AMM) uses the formula
\[ x \cdot y = k \]
and the price is given by
\[ -\frac{dy}{dx} = \frac{y}{x}. \]
This approach was formalised in the Uniswap V2 Protocol~\cite{uniswap_v2_core}.

In QubitSwap, we incorporate an oracle price \( p \) and a mixing factor \( z \), where \( z \in [0, 1] \). Similar approaches using oracle prices have been implemented in platforms like GMX to achieve reduced slippage in decentralised trading~\cite{gmx2023}. The modified pricing equation is
\[ -\frac{dy}{dx} = (1 - z) \frac{y}{x} + z p. \]
This equation represents a weighted blend of the internal pool price and the external oracle price, with interpretations:
- When \( z = 0 \), it reduces to the classic AMM behaviour: \( -\frac{dy}{dx} = \frac{y}{x} \).
- When \( z = 1 \), the price fully follows the oracle: \( -\frac{dy}{dx} = p \).
- For \( 0 < z < 1 \), it is a weighted average of the two.

\section{Derivation of the Differential Equation}
Rearranging the pricing equation, we obtain
\[ \frac{dy}{dx} + (1 - z) \frac{y}{x} = -z p. \]
This is a first-order linear ordinary differential equation (ODE) in \( y \) with respect to \( x \).

\section{Solving the Differential Equation}
To solve the ODE, we use the integrating factor method. The integrating factor is
\[ \text{IF} = x^{1 - z}. \]
Multiplying both sides of the ODE by the integrating factor, we get
\[ \frac{d}{dx} \left( y \cdot x^{1 - z} \right) = -z p x^{1 - z}. \]
Integrating both sides with respect to \( x \):
\[ y x^{1 - z} = -z p \int x^{1 - z} \, dx + C, \]
which yields
\[ y x^{1 - z} = -z p \cdot \frac{x^{2 - z}}{2 - z} + k. \]
Solving for \( y \), we obtain
\[ y = -\frac{z p x}{2 - z} + k x^{z - 1}. \]
Thus, the key reserve function is
\[ y(x) = k x^{z - 1} - \frac{z p x}{2 - z}, \]
where \( k \) is a constant determined by initial conditions.

The differential formula, representing the rate of change of \( y \) with respect to \( x \), is
\[ \frac{dy}{dx} = k (z - 1) x^{z - 2} - \frac{z p}{2 - z}. \]

\begin{figure}[h]
  \centering
  \includegraphics[width=0.7\textwidth]{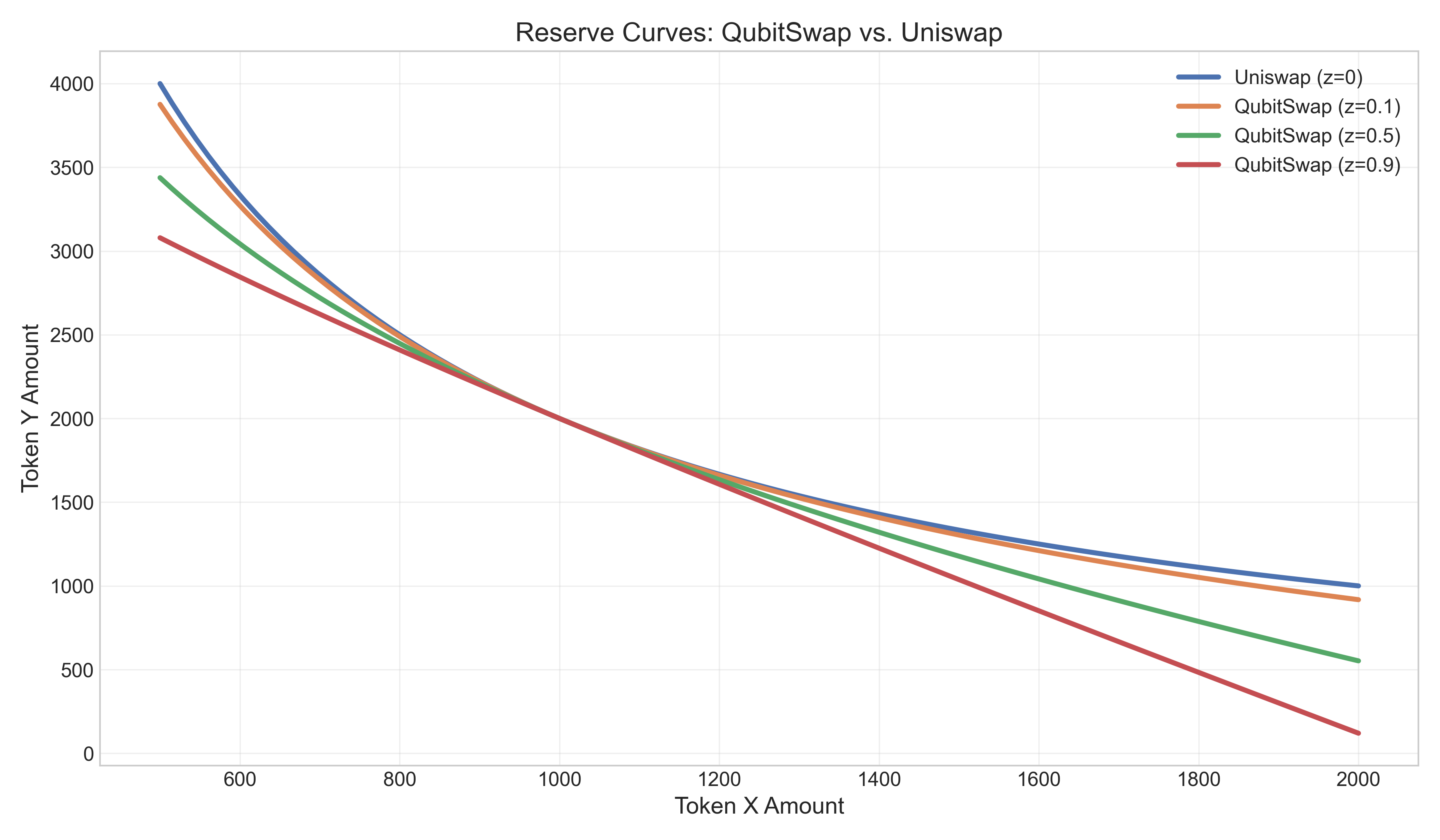}
  \caption{Comparison between classic AMM curve (z=0) and QubitSwap curves with different z values.}
  \label{fig:reserve_curve_comparison}
\end{figure}

\section{Insights and Properties}
The reserve function exhibits distinct behaviours at the extremes:
- When \( z = 0 \):
  \[ y = k x^{-1}, \]
  which matches the classic AMM constant product model.
- When \( z = 1 \):
  \[ y = -p x + k, \]
  indicating a linear relationship driven by the oracle price.

For \( z > 0 \), the reserve function is asymmetric, unlike the symmetric hyperbola of the classic AMM. This asymmetry, combined with the blending of the oracle price, reduces sensitivity to rapid price swings, benefiting liquidity providers.

\section{Constant Domains}
The parameters and variables are defined within the following domains:
- Mixing factor: \( z \in [0, 1] \)
- Reserves: \( x, y > 0 \)
- Oracle price: \( p > 0 \)
- Constant \( k \): Adjusted to ensure \( y > 0 \)

\section{Impermanent Loss (IL) Analysis}
Impermanent loss (IL) is the value difference between holding assets versus providing liquidity due to price changes. In standard AMMs, for a price ratio \( r = \frac{p_1}{p_0} \), the IL is given by
\[ \text{IL} = 2 \sqrt{r} - r - 1. \]

Recent research by Haddad et al. provides a formal analysis of the conditions under which liquidity providers experience impermanent loss in decentralised exchanges~\cite{haddad2024}.

\begin{figure}[h]
  \centering
  \includegraphics[width=0.7\textwidth]{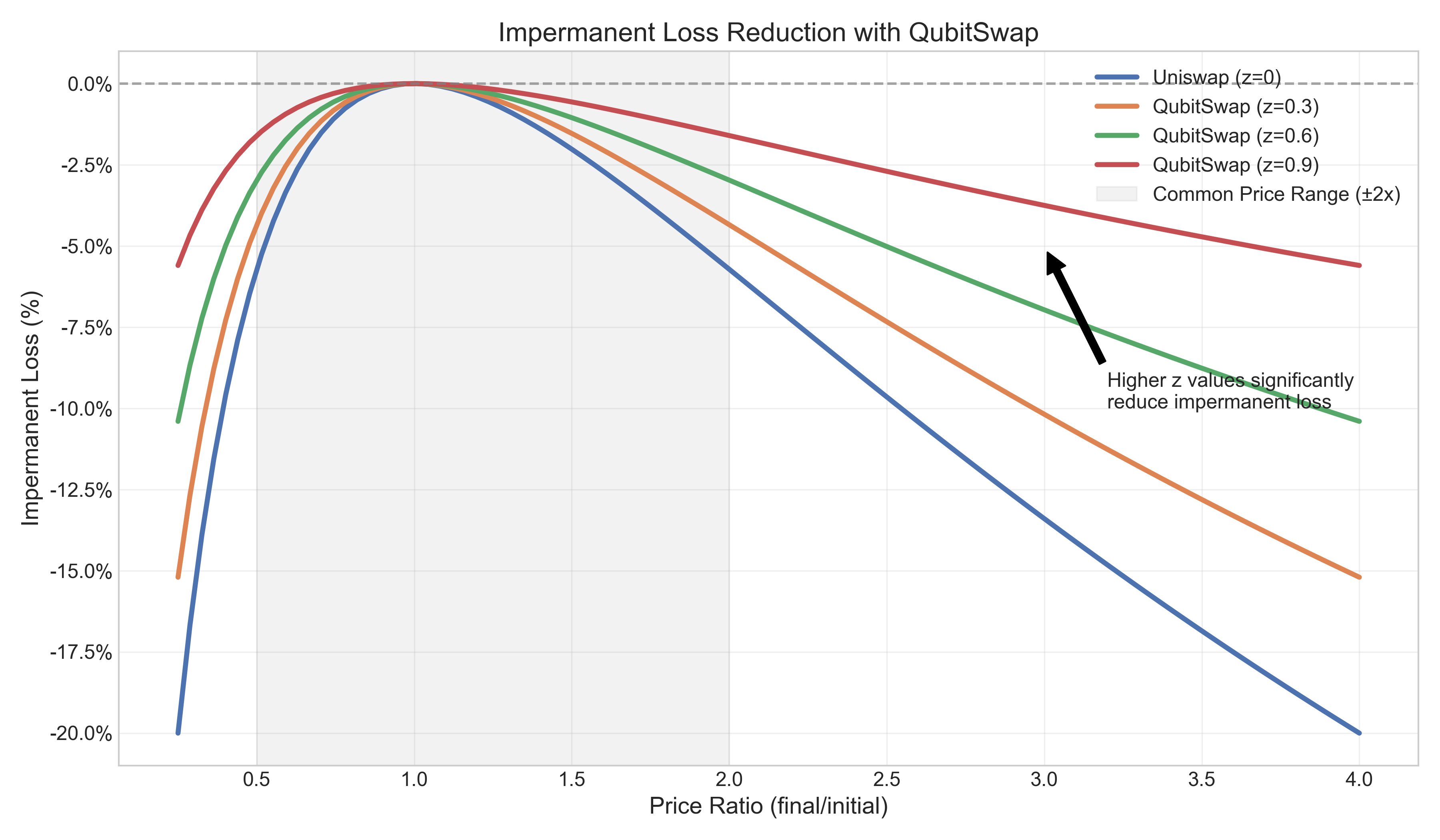}
  \caption{Impermanent loss comparison between classic AMM (z=0) and QubitSwap (z=0.3, 0.6, 0.9).}
  \label{fig:impermanent_loss_comparison}
\end{figure}

In QubitSwap, the reserve adjustment is derived as
\[ x_1 = x_0 \left( \frac{p_0}{p_1} \right)^{\frac{1}{2 - z}}, \]
\[ y_1 = p_1 x_1. \]
The pool value is
\[ V_{\text{pool}} = x_1 + p_1 y_1 = 2 x_0 \left( \frac{p_0}{p_1} \right)^{\frac{1}{2 - z}}, \]
and the holding value is
\[ V_{\text{hold}} = x_0 + p_1 y_0 = x_0 \left( 1 + \frac{p_0}{p_1} \right). \]
Thus, the IL for QubitSwap is
\[ \text{IL} = \frac{V_{\text{hold}} - V_{\text{pool}}}{V_{\text{hold}}} = 1 + \frac{p_0}{p_1} - 2 \left( \frac{p_0}{p_1} \right)^{\frac{1}{2 - z}}. \]

For \( z > 0 \), this IL is reduced compared to the standard AMM (\( z = 0 \)).

\section{Slippage Analysis}
Slippage is the difference between the expected and executed trade price due to reserve impact. Using a Taylor expansion of \( y(x) \) for a trade size \( \Delta x \):
\[ \Delta y \approx \frac{dy}{dx} \Delta x + \frac{1}{2} \frac{d^2 y}{dx^2} (\Delta x)^2. \]
The average trade price is
\[ p_{\text{avg}} = -\frac{\Delta y}{\Delta x} \approx -\frac{dy}{dx} - \frac{1}{2} \frac{d^2 y}{dx^2} \Delta x, \]
with the spot price \( p_0 = -\frac{dy}{dx} \). Thus, slippage is
\[ \text{slippage} = p_{\text{avg}} - p_0 \approx -\frac{1}{2} \frac{d^2 y}{dx^2} \Delta x. \]
Computing the second derivative:
\[ \frac{dy}{dx} = k (z - 1) x^{z - 2} - \frac{z p}{2 - z}, \]
\[ \frac{d^2 y}{dx^2} = k (z - 1)(z - 2) x^{z - 3}. \]
Substituting:
\[ \text{slippage} \approx -\frac{1}{2} k (z - 1)(z - 2) x^{z - 3} \Delta x. \]
Relating \( k (z - 1) x^{z - 2} = \frac{dy}{dx} + \frac{z p}{2 - z} \), we can simplify to:
\[ \text{slippage} \approx -\frac{\Delta x (z - 2)}{x} \left( \frac{dy}{dx} + \frac{z p}{2 - z} \right). \]
Slippage scales with \( \Delta x \), \( x \), \( z \), and \( p \), decreasing as \( z \) nears 1.

\subsection{Concentration Effect of \( z \)}
The parameter \( z \) in QubitSwap plays a pivotal role in controlling slippage by balancing the influence of internal pool dynamics and an external oracle price \( p \). As \( z \) approaches 1, slippage diminishes significantly, creating what we term the \textbf{concentration effect}. This effect has conceptual similarities to the concentrated liquidity mechanism introduced in Uniswap V3~\cite{uniswap_v3_core}, though achieved through different mathematical approaches. The concentration effect describes how the pool behaves as if it has a high concentration of liquidity centered around the oracle price, reducing the price impact of trades and allowing larger trades to be executed with minimal slippage.

To understand this, consider the role of \( z \):
- When \( z = 0 \), the pool operates like a traditional automated market maker (AMM), relying solely on internal reserves. Here, large trades significantly deplete reserves, leading to substantial price movements and high slippage.
- When \( z = 1 \), the pool fully adopts the oracle price, making the price insensitive to reserve changes (\( \frac{dy}{dx} = -p \), and \( \frac{d^2 y}{dx^2} = 0 \)), resulting in zero slippage.
- For \( z \in (0,1) \), the pool blends these dynamics, with increasing \( z \) shifting reliance toward the oracle price.

The slippage formula provides insight into this behaviour. Using the second-derivative form for simplicity:

\[ \text{slippage} \approx -\frac{1}{2} k (z - 1)(z - 2) x^{z - 3} \Delta x \]

For \( z \in (0,1) \), both \( (z - 1) \) and \( (z - 2) \) are negative, so \( (z - 1)(z - 2) > 0 \), and since \( k > 0 \), \( x > 0 \), and typically \( \Delta x > 0 \) (for a buy trade), the slippage is negative in this form. However, in decentralised exchange (DEX) contexts, slippage is conventionally the additional cost to the trader (\( p_{\text{avg}} - p_0 > 0 \) for buying), so we adjust the definition to:

\[ \text{slippage} \approx \frac{1}{2} k (z - 1)(z - 2) x^{z - 3} \Delta x \]

Now, as \( z \to 1^- \):
- \( (z - 1) \to 0^- \), so \( (z - 1)(z - 2) \to 0 \cdot (1 - 2) = 0 \).
- The slippage approaches zero, indicating that the price impact vanishes.

This reduction reflects the pool's increasing stability. To quantify this, consider a normalised example with initial reserves \( x_0 = 1 \), oracle price \( p = 1 \), and \( y_0 = 1 \). From \( y(x) = k x^{z - 1} - \frac{z p x}{2 - z} \), at \( x = 1 \):

\[ y(1) = k - \frac{z}{2 - z} = 1 \quad \Rightarrow \quad k = 1 + \frac{z}{2 - z} \]

Thus, slippage becomes:

\[ \text{slippage} \approx \frac{1}{2} \left( 1 + \frac{z}{2 - z} \right) (z - 1)(z - 2) \Delta x \]

- At \( z = 0.1 \):
  \[ k = 1 + \frac{0.1}{1.9} = \frac{20}{19}, \quad (z - 1)(z - 2) = (-0.9)(-1.9) = 1.71 \]
  \[ \text{slippage} \approx \frac{1}{2} \cdot \frac{20}{19} \cdot 1.71 \cdot \Delta x \approx 0.9 \Delta x \]

- At \( z = 0.9 \):
  \[ k = 1 + \frac{0.9}{1.1} = \frac{20}{11}, \quad (z - 1)(z - 2) = (-0.1)(-1.1) = 0.11 \]
  \[ \text{slippage} \approx \frac{1}{2} \cdot \frac{20}{11} \cdot 0.11 \cdot \Delta x = 0.1 \Delta x \]

As \( z \) increases from 0.1 to 0.9, slippage drops from \( 0.9 \Delta x \) to \( 0.1 \Delta x \), a ninefold reduction. As \( z \to 1 \), slippage approaches zero, mimicking the behaviour of a pool with infinite liquidity concentrated at \( p \). This \textbf{concentration effect} enhances QubitSwap's efficiency, making it more resilient to large trades compared to traditional AMMs, where slippage grows with trade size due to reserve depletion.

In essence, the concentration effect arises because higher \( z \) values reduce the pool's sensitivity to reserve changes, effectively concentrating liquidity around the oracle price and minimising price impact, much like concentrated liquidity mechanisms in other AMMs (e.g., Uniswap V3), but achieved here through the \( z \)-parameterised pricing model.

\begin{figure}[h]
  \centering
  \includegraphics[width=0.7\textwidth]{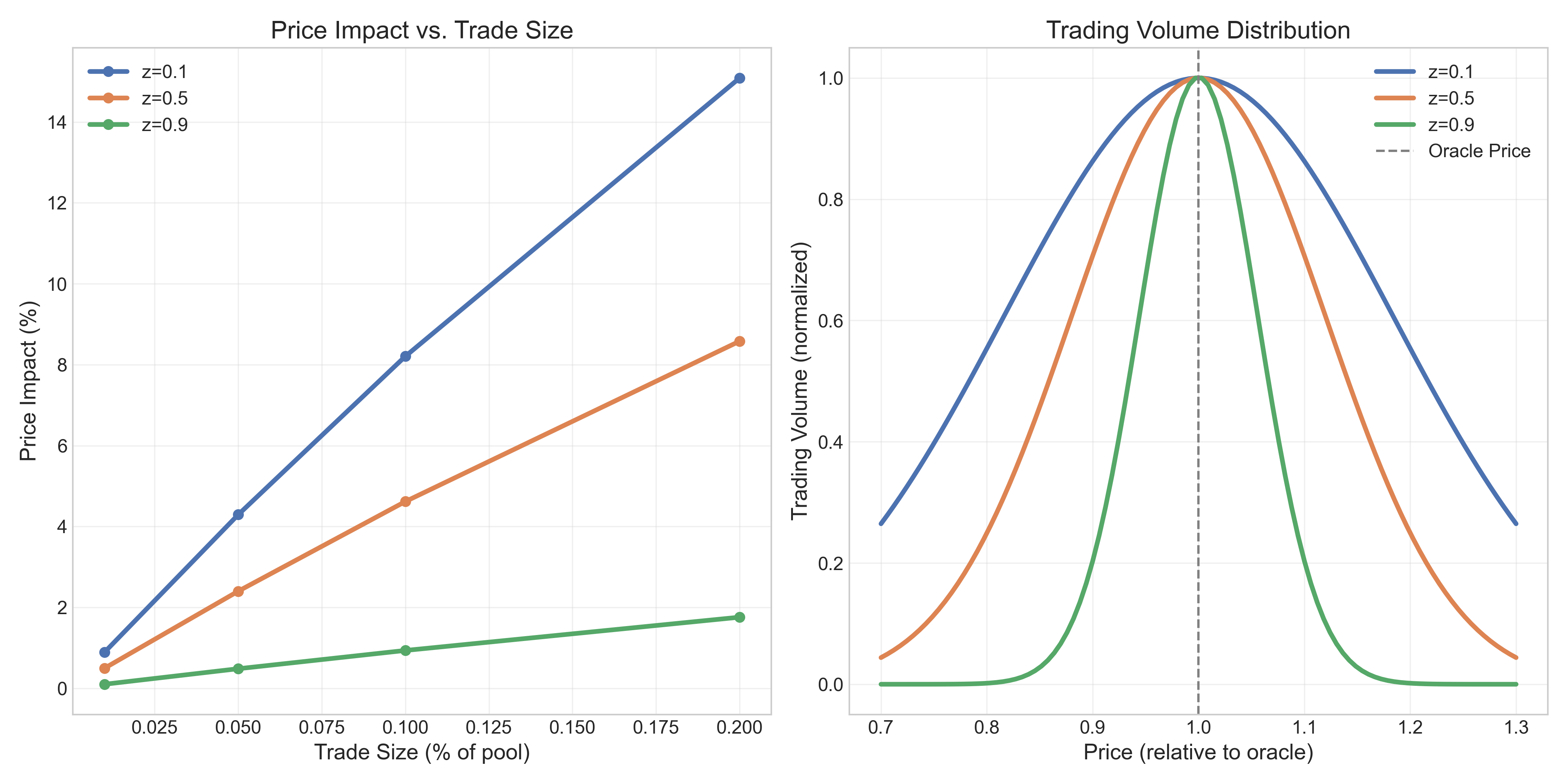}
  \caption{Trading simulations between classic AMM (z=0) and QubitSwap (z=0.1, 0.5, 0.9).}
  \label{fig:trading_simulation}
\end{figure}

\section{Key Takeaways}
- \textbf{Reserve Function}: 
  \[ y(x) = k x^{z - 1} - \frac{z p x}{2 - z} \]
- \textbf{Price Equation}: 
  \[ -\frac{dy}{dx} = (1 - z) \frac{y}{x} + z p \]
- \textbf{Differential Formula}: 
  \[ \frac{dy}{dx} = k (z - 1) x^{z - 2} - \frac{z p}{2 - z} \]
- \textbf{Slippage Formula}: 
  \[ \text{slippage} \approx -\frac{\Delta x (z - 2)}{x} \left( \frac{dy}{dx} + \frac{z p}{2 - z} \right) \]
- QubitSwap reduces IL and slippage compared to traditional AMMs by integrating an oracle price, enhancing efficiency for traders and liquidity providers.

\section{Conclusion}
QubitSwap introduces a new simple approach to decentralised exchanges (DEXs) by combining external oracle pricing with internal pool dynamics. This hybrid design rigorously tackles the issues of impermanent loss and slippage common in traditional automated market makers (AMMs). Using a parameter \( z \), the model adjusts how much it relies on these price sources, creating a flexible pricing system. Our analysis shows it reduces financial risk for liquidity providers and offers more predictable pricing for traders, even when markets are volatile. QubitSwap adds to the effort to build more efficient decentralised trading systems and opens the door for further improvements through future research.

\section{Disclaimer}
This paper is for general information purposes only. It does not constitute investment advice or a recommendation or solicitation to buy or sell any investment, and should not be used as the basis for any investment decision or for accounting, legal, or tax advice. The opinions expressed are those of the authors and do not necessarily reflect the views of any organisation or entity.
\end{document}